\documentclass[12pt]{article}
\usepackage{graphicx,amsmath}
\usepackage[pagebackref=false, colorlinks, linkcolor=black, citecolor=black]{hyperref}
\def\nn{\nonumber}
\def\pa{\partial}
\def\ie{\textit{i.e.}}

\newcommand{\eg}{{\it e.g.,}\ }
\begin{document}
\begin{titlepage}
\begin{center}
\vskip 2 cm
{\LARGE \bf  Minimal independent couplings\\  \vskip 0.25 cm at order $\alpha'^2$  
 }\\
\vskip 1.25 cm
 Mohammad R. Garousi\footnote{garousi@um.ac.ir} and  Hamid Razaghian\footnote{razaghian.hamid@gmail.com}

\vskip 1 cm
{{\it Department of Physics, Faculty of Science, Ferdowsi University of Mashhad\\}{\it P.O. Box 1436, Mashhad, Iran}\\}
\vskip .1 cm
 \end{center}
\begin{abstract}
Using field redefinitions  and  Bianchi identities on the general form of the effective action  for metric, $B$-field and dilaton, we have found that the minimum number of  independent couplings at order $\alpha'^2$  is 60. We write these couplings in two different schemes in the string frame. In the first  scheme, each coupling does not include terms with more than two derivatives  and  it does not include structures $R,\,R_{\mu\nu},\,\nabla_\mu H^{\mu\alpha\beta}$, $ \nabla_\mu\nabla^\mu\Phi$. In this scheme, 20  couplings which are the minimum number of  couplings for metric and $B$-field,  include   dilaton trivially as the overall factor of $e^{-2\Phi}$, and all other couplings  include derivatives of dilaton.  In the second scheme, the dilaton appears in all 60  coupling only as the overall factor of $e^{-2\Phi}$ . In this scheme,  20 of the couplings are exactly the same as those  in the previous scheme. 

\end{abstract}
\end{titlepage}
\tableofcontents

\section{Introduction} \label{intro}
String theory is a quantum theory of gravity   with  a finite number of massless fields and a  tower of infinite number of  massive fields reflecting the stringy nature of the gravity.
An efficient way to study different phenomena in this theory is to use an effective action  which includes     only  massless fields and their derivatives \cite{Howe:1983sra,Witten:1995ex}. 
The effective action  has a double expansions. The genus expansion which includes  the  classical and a tower of quantum corrections, and the   stringy expansion which is an expansion   in powers of the Regge slope parameter $\alpha'$. The latter expansion for metric yields  the Einstein gravity and the stringy corrections which are    quadratic and higher orders in curvature.
The main challenge thus is to explore different techniques   to find the effective action that incorporates  all such corrections, including  non-perturbative effects \cite{Green:1997tv}.
In the bosonic and in the heterotic string theories, the higher derivative couplings  begin at order $\alpha'$, whereas, in type II superstring theory, they   begin at order $\alpha'^3$.

There are various techniques  in the string theory for finding these higher derivative couplings:
S-matrix element approach \cite{Scherk:1974mc,Yoneya:1974jg}, sigma-model approach \cite{Callan:1985ia,Fradkin:1984pq,Fradkin:1985fq},  supersymmetry approach \cite{Gates:1986dm,Gates:1985wh,Bergshoeff:1986wc,Bergshoeff:1989de}, double field theory  approach  \cite{Siegel:1993xq,Hull:2009mi,Hohm:2010jy},  and duality approach \cite{Ferrara:1989bc,Font:1990gx,Green:1997tv,Garousi:2017fbe}.
In the duality approach, the  consistency of the effective actions with T- and S-duality transformations are imposed to find the higher derivative couplings \cite{Green:1997tv,Garousi:2017fbe}.
 In particular, it has been speculated in   \cite{Razaghian:2017okr} that the consistency of the effective actions at any order of $\alpha'$ with the T-duality transformations may fix both the effective actions and the corrections to the Buscher rules   \cite{Buscher:1987sk,Buscher:1987qj}. It has been shown explicitly  in \cite{Garousi:2019wgz} that the T-duality constraint fixes   the effective action and the corrections to the Buscher rules at order $\alpha'$, up to an overall factor.

In using the above techniques for finding the effective actions at the higher-derivative orders in the string theory, one needs the most general gauge invariant  and minimal independent couplings at each order of $\alpha'$. To find such couplings, one needs to impose various Bianchi identities   and use  field redefinitions freedom \cite{Gross:1986iv,Tseytlin:1986ti,Deser:1986xr}. In the literature,  the  Bianchi identities are first  imposed to find the minimum number of couplings at  each order of $\alpha'$, up to some total derivative terms and field redefinitions. The parameters in the resulting action are then either unambiguous which are not changed under field redefinition, or ambiguous which are changed under the field redefinitions. Some combinations of the latter parameters, however, remain invariant under the field redefinitions  \cite{Metsaev:1987zx}.  This allows one to separate the ambiguous parameters to essential parameters which are fixed by \eg S-matrix calculations  \cite{Metsaev:1987zx,Metsaev:1986yb}, and some remaining arbitrary parameters. Depending on which set of parameters are chosen as essential parameters, one has different schemes. To find the minimum number of independent couplings, one sets all the arbitrary parameters to zero.    This method has been used  to find the 8 independent couplings for gravity, $B$-field and dilaton at order $\alpha'$ in  \cite{Metsaev:1987zx}, the 7 independent couplings for gravity and dilaton  at order $\alpha'^2$  in \cite{Bento1989,Bento:1989ir,Bento1990,Bento:1990jc}   and 20 independent couplings for gravity and B-field  at order $\alpha'^2$ in  \cite{Jones:1988hk}\footnote{The authors in  \cite{Jones:1988hk} reported that there are 21 independent couplings at order $\alpha'^2$. We think there should be  a typo in writing the number of independent $H^6$ terms in \cite{Jones:1988hk}.  They have written 8 independent terms with structure $H^6$, whereas, we have found that there are only 7 such  terms. That indicates that there should be  20 independent couplings for gravity and B-field, .}. 

One may impose  the Bianchi identities,  remove  the boundary terms and  use the  field redefinition freedom at the same time. That is, one may first write  all  gauge invariant couplings at each order of $\alpha'$ and then impose the above freedoms to reduce the couplings to the minimal couplings. The parameters in the gauge invariant  action are then either unambiguous or ambiguous depending on whether or not they are changed under these freedoms.   Some combinations of the ambiguous  parameters, however, remain invariant.  This allows one to separate the ambiguous parameters to essential parameters which may be found  by  S-matrix calculations, and some arbitrary parameters. Again, depending on which set of parameters are choosing as essential parameters, one has different schemes. The minimum number of independent couplings are found in the schemes that all the arbitrary parameters are set to zero.  We find that this latter method is more convenient to find the independent couplings systematically,  using the Mathematica packages  like  "xAct" \cite{Nutma:2013zea}. In particular, to impose the Bianchi identities we write the curvatures and the covariant derivatives in terms of metric and its derivatives. Then we choose the local inertial frame in which the first derivative of metric is zero. In this frame the Bianchi identities   are all satisfied automatically.  Using this   method, we are going to find the   minimal  independent couplings  for gravity, dilaton and B-field at order $\alpha'^2$. We find that there are 60 parameters in the minimal couplings. We write them in two different schemes. Both schemes have the same 20 couplings between gravity and $B$-field. In one scheme the other 40 couplings include derivatives of dilaton, and in the other scheme the 40 couplings does not include the derivative of the dilaton. The 20 common couplings in both schemes are the minimal couplings when dilaton is constant.


The outline of the paper is as follows: In section 2, we write the most general gauge invariant couplings involving, metric, dilaton and B-field at order $\alpha'$. There are 41 such terms. Then we add to them the most general boundary terms and field redefinitions with arbitrary parameters. Writing them in the local inertial frame,  we then use the arbitrary parameters in the total derivative terms and in the field redefinitions  to reduce the 41 couplings to 8 independent couplings that are known in the literature. We write them in two different schemes. In one scheme, there is no  term in which fields have more than two derivatives and there is no term  involving $R,R_{\mu\nu},\nabla_\mu H^{\mu\alpha\beta}$, $ \nabla_\mu\nabla^\mu\Phi$.  More specificity, we write the couplings into two separate parts. One part which has 4 couplings, does not include derivatives of dilaton and it is the same as the set of minimal couplings when dilaton is constant. The other part includes derivatives of dilaton. In the second scheme, we again write the couplings  into two parts, one part is the same as the minimal couplings when dilaton is constant, and the other part includes some other couplings in which dilaton appears trivially. In section 3, we extend the calculations to the order $\alpha'^2$. We found that the most general action at this order has 705 couplings, however, adding the total derivative terms and  field redefinitions to them  with arbitrary parameters, and writing the result in the local frame,  we find that the arbitrary parameters can be used to reduce the couplings to the minimum number of   couplings  which  is 60. We write them in two different   schemes as in the section 2.  Each scheme has 20 common couplings which are the   minimal couplings when dilaton is constant,  and 40 other couplings. These couplings all include derivatives of dilaton in one scheme, whereas, in the other scheme the dilaton appears trivially. 

\section{Minimal couplings at order  $\alpha'$}\label{sec.2}

The effective action of string theory has a double expansions. One expansion is the genus expansion which includes  the  classical sphere-level and a tower of quantum effects. The other one is a stringy expansion which is an expansion in terms of higher-derivative couplings. The number of derivatives in each  coupling can be accounted by the order of $\alpha'$. The sphere-level effective action  has the following    power  series of $\alpha'$ in the string frame:
\begin{align}
S_{\rm eff}&=\sum^\infty_{n=0}\alpha'^nS_n=S_0+\alpha' S_1+\alpha'^2 S_2+\cdots\ ; \quad S_n=\int d^D x\sqrt{-g} e^{-2\Phi}\mathcal{L}_n\label{eq.13021218}
\end{align}
The effective action must be invariant under the coordinate transformations and under the $B$-field   gauge transformations. So the metric $g$, the  antisymmetric  $B$-field and dilaton       must  appear in the Lagrangian $ \mathcal{L}_n$  trough their field strengths and their covariant derivatives, \eg
the Lagrangian at the leading order of $\alpha'$ is\footnote{Throughout in this paper, we use the conventions
\begin{align}
&R^\lambda{}_{\mu\nu\rho}=\pa_\nu \Gamma^\lambda_{\mu\rho}-\cdots, \qquad R_{\mu\nu}=R^\lambda{}_{\mu\lambda\rho}\nn\\
&H_{\lambda\mu\nu}=3\partial_{[\lambda}B_{\mu\nu]}, \qquad\qquad\nabla_\mu A^\lambda=\partial_\mu A^\lambda+\Gamma^\lambda_{\mu\nu}A^\nu,\nn\\
&T_{[\mu_1\ldots\mu_n]}=\frac{1}{n!}(T_{\mu_1\ldots\mu_n}+\cdots).
\end{align}}:
\begin{align}
\mathcal{L}_0&= R-\frac{1}{12}H_{\alpha\beta\gamma}H^{\alpha\beta\gamma}+4\nabla_\alpha\Phi\nabla^\alpha\Phi
\end{align}
The  higher-derivative field redefinitions   and Bianchi identity can not change the form of this action.

Using the Bianchi identities, it has been shown in  \cite{Metsaev:1987zx} that, up to some boundary terms,  the Lagrangian $\mathcal{L}_1$  has 20   couplings, each with an arbitrary parameter.  3 of these parameters  are unambiguous because they are not changed under field redefinitions, and all others are ambiguous. The field redefinition freedom then has been used to show that only 5 couplings among the ambiguous couplings are essential and all others are arbitrary. To find the minimal independent couplings, one sets the arbitrary parameters to zero \cite{Metsaev:1987zx}. In this section, we are going to re-derive the 8 independent  couplings by using a systematic method for  using total derivative terms, applying  the field redefinitions and the Bianchi identities, that can easily be extended to the higher order couplings. 

Using the package ``xAct``, one finds the most general gauge invariant Lagrangian at order $\alpha'$ has the following couplings:
\begin{align}
 \mathcal{L}_1=&   B_1  R_{\alpha \beta \gamma \delta} R^{\alpha \beta \gamma \delta} + B_2  R_{\alpha \gamma \beta \delta} R^{\alpha \beta \gamma \delta}  + B_3  R_{\alpha \beta} R^{\alpha \beta} + B_4  R^2+ B_5  \nabla_{\beta}\nabla_{\alpha}R^{\alpha \beta} + B_6  \nabla_{\alpha}\nabla^{\alpha}R \nn\\
& + B_7  R^{\alpha \beta} \nabla_{\beta}\nabla_{\alpha}\Phi 
+ B_8  R_{\alpha \beta} \nabla^{\alpha}\Phi \nabla^{\beta}\Phi 
+ B_9  R \nabla_{\alpha}\nabla^{\alpha}\Phi 
+ B_{10}  R \nabla_{\alpha}\Phi \nabla^{\alpha}\Phi  
+ B_{11}  \nabla^{\alpha}\Phi \nabla_{\beta}R_{\alpha}{}^{\beta} \nn\\
& + B_{12}  \nabla_{\alpha}\Phi \nabla^{\alpha}R 
+ B_{13}  \nabla_{\beta}\nabla^{\beta}\nabla_{\alpha}\nabla^{\alpha}\Phi  
+ B_{14}  \nabla_{\beta}\nabla_{\alpha}\nabla^{\beta}\nabla^{\alpha}\Phi  
+ B_{15} \nabla_{\alpha}\nabla_{\beta}\nabla^{\beta}\Phi \nabla^{\alpha}\Phi\nn\\
& + B_{16}  \nabla^{\alpha}\Phi \nabla_{\beta}\nabla^{\beta}\nabla_{\alpha}\Phi 
+ B_{17}  \nabla_{\beta}\nabla_{\alpha}\Phi \nabla^{\beta}\nabla^{\alpha}\Phi 
+ B_{18}  \nabla_{\alpha}\nabla^{\alpha}\Phi \nabla_{\beta}\nabla^{\beta}\Phi \nn\\
& + B_{19}  \nabla^{\alpha}\Phi \nabla_{\beta}\nabla_{\alpha}\Phi \nabla^{\beta}\Phi 
+ B_{20}  \nabla_{\alpha}\Phi \nabla^{\alpha}\Phi \nabla_{\beta}\nabla^{\beta}\Phi 
+ B_{21}  \nabla_{\alpha}\Phi \nabla^{\alpha}\Phi \nabla_{\beta}\Phi \nabla^{\beta}\Phi\nn\\
&+ B_{22}  H_{\alpha}{}^{\delta \epsilon} H^{\alpha \beta \gamma} H_{\beta \delta}{}^{\zeta} H_{\gamma \epsilon \zeta}
+ B_{23}  H_{\alpha \beta}{}^{\delta} H^{\alpha \beta \gamma} H_{\gamma}{}^{\epsilon \zeta} H_{\delta \epsilon \zeta} 
+ B_{24}  H_{\alpha \beta \gamma} H^{\alpha \beta \gamma} H_{\delta \epsilon \zeta} H^{\delta \epsilon \zeta}  \nn\\
&+ B_{25}  H^{\alpha \beta \gamma} \nabla_{\delta}\nabla^{\delta}H_{\alpha \beta \gamma} 
+ B_{26}  H^{\alpha \beta \gamma} \nabla_{\delta}\nabla_{\gamma}H_{\alpha \beta}{}^{\delta}  
+ B_{27}  H^{\alpha \beta \gamma} \nabla_{\gamma}\nabla_{\delta}H_{\alpha \beta}{}^{\delta} 
+ B_{28}  \nabla_{\delta}H_{\alpha \beta \gamma} \nabla^{\delta}H^{\alpha \beta \gamma}\nn\\
&+ B_{29}  \nabla_{\alpha}H^{\alpha \beta \gamma} \nabla_{\delta}H_{\beta \gamma}{}^{\delta} 
+ B_{30}  \nabla_{\gamma}H_{\alpha \beta \delta} \nabla^{\delta}H^{\alpha \beta \gamma}
+ B_{31}  H_{\alpha}{}^{\delta \epsilon} H^{\alpha \beta \gamma} R_{\beta \gamma \delta \epsilon}
+ B_{32}  H_{\alpha}{}^{\delta \epsilon} H^{\alpha \beta \gamma} R_{\beta \delta \gamma \epsilon}\nn\\
&+ B_{33}  H_{\alpha}{}^{\gamma \delta} H_{\beta \gamma \delta} R^{\alpha \beta} 
+ B_{34}  H_{\alpha \beta \gamma} H^{\alpha \beta \gamma} R 
+ B_{35}  H^{\beta \gamma \delta} \nabla_{\alpha}H_{\beta \gamma \delta} \nabla^{\alpha}\Phi 
+ B_{36}  H^{\beta \gamma \delta} \nabla^{\alpha}\Phi \nabla_{\delta}H_{\alpha \beta \gamma}  \nn\\
&+ B_{37}  H_{\alpha}{}^{\beta \gamma} \nabla^{\alpha}\Phi \nabla_{\delta}H_{\beta \gamma}{}^{\delta} 
+ B_{38}  H_{\beta \gamma \delta} H^{\beta \gamma \delta} \nabla_{\alpha}\nabla^{\alpha}\Phi  
+ B_{39}  H_{\alpha}{}^{\gamma \delta} H_{\beta \gamma \delta} \nabla^{\beta}\nabla^{\alpha}\Phi  \nn\\
&+ B_{40}  H_{\beta \gamma \delta} H^{\beta \gamma \delta} \nabla_{\alpha}\Phi \nabla^{\alpha}\Phi+ B_{41}  H_{\alpha}{}^{\gamma \delta} H_{\beta \gamma \delta} \nabla^{\alpha}\Phi \nabla^{\beta}\Phi 
\label{eq.04301918}
\end{align}
where $B_1,\cdots, B_{41}$ are some parameters. The above couplings are  not all independent. Some of them are related by total derivative terms, some of them are related by field redefinitions, and some others are related by various Bianchi identities.

To remove the total derivative terms from the above couplings, we consider the most general total derivative terms at order $\alpha'$ which has the following structure:
\begin{align}
\alpha'\int d^Dx \sqrt{-g}e^{-2\Phi} \mathcal{J}_1=\alpha'\int d^Dx\sqrt{-g} \nabla_\alpha (e^{-2\Phi}J_1^\alpha) \label{eq.04301919}
\end{align}
where the vector $J^\alpha$ is   all possible covariant and gauge invariant  terms at three-derivative level, \ie, 
\begin{align}
J_1^\alpha= &J_1 \nabla_{\beta}R^{\alpha \beta}  + J_2 \nabla^{\alpha}R   + J_3 R^{\alpha}{}_{\beta} \nabla^{\beta}\Phi + J_4 R \nabla^{\alpha}\Phi   + J_5\nabla_{\beta}\nabla^{\beta}\nabla^{\alpha}\Phi  + J_6 \nabla^{\alpha}\nabla_{\beta}\nabla^{\beta}\Phi \nn\\
&   + J_7 \nabla_{\beta}\nabla^{\alpha}\Phi \nabla^{\beta}\Phi  + J_8 \nabla^{\alpha}\Phi \nabla_{\beta}\nabla^{\beta}\Phi+ J_9 \nabla^{\alpha}\Phi \nabla_{\beta}\Phi \nabla^{\beta}\Phi +J_{10} H^{\beta \gamma \delta} \nabla^{\alpha}H_{\beta \gamma \delta}   \nn\\
&  + J_{11} H^{\beta \gamma \delta} \nabla_{\delta}H^{\alpha}{}_{\beta \gamma}  + J_{12} H^{\alpha \beta \gamma} \nabla_{\delta}H_{\beta \gamma}{}^{\delta}  + J_{13} H_{\beta \gamma \delta} H^{\beta \gamma \delta} \nabla^{\alpha}\Phi   + J_{14} H^{\alpha \gamma \delta} H_{\beta \gamma \delta} \nabla^{\beta}\Phi \label{eq.04301833}
\end{align}
where the coefficients  $J_1,\cdots, J_{14}$ are arbitrary parameters. Inserting this into   \eqref{eq.04301919}, one finds
\begin{align}
 \mathcal{J}_1=& J_1 \nabla_{\beta}\nabla_{\alpha}R^{\alpha \beta} +J_2 \nabla_{\alpha}\nabla^{\alpha}R + J_4 R \nabla_{\alpha}\nabla^{\alpha}\Phi + (-2 J_2 + J_4) \nabla_{\alpha}\Phi \nabla^{\alpha}R  \nonumber \\ 
& + (-2 J_{10}+ 2 J_{13}) H^{\beta \gamma \delta} \nabla_{\alpha}H_{\beta \gamma \delta} \nabla^{\alpha}\Phi - 2 J_{13} H_{\beta \gamma \delta} H^{\beta \gamma \delta} \nabla_{\alpha}\Phi \nabla^{\alpha}\Phi - 2 J_4 R \nabla_{\alpha}\Phi \nabla^{\alpha}\Phi  \nonumber \\ 
& + (-2 J_6 + J_8) \nabla_{\alpha}\nabla_{\beta}\nabla^{\beta}\Phi \nabla^{\alpha}\Phi+ (-2 J_1 + J_3) \nabla^{\alpha}\Phi \nabla_{\beta}R_{\alpha}{}^{\beta} + J_3 R^{\alpha \beta} \nabla_{\beta}\nabla_{\alpha}\Phi  \nonumber \\ 
& + J_8 \nabla_{\alpha}\nabla^{\alpha}\Phi \nabla_{\beta}\nabla^{\beta}\Phi+ (-2 J_8 + J_9) \nabla_{\alpha}\Phi \nabla^{\alpha}\Phi \nabla_{\beta}\nabla^{\beta}\Phi + (-2 J_5 + J_7) \nabla^{\alpha}\Phi \nabla_{\beta}\nabla^{\beta}\nabla_{\alpha}\Phi \nonumber \\ 
& + J_6 \nabla_{\beta}\nabla^{\beta}\nabla_{\alpha}\nabla^{\alpha}\Phi - 2 J_{14} H_{\alpha}{}^{\gamma \delta} H_{\beta \gamma \delta} \nabla^{\alpha}\Phi \nabla^{\beta}\Phi - 2 J_3 R_{\alpha \beta} \nabla^{\alpha}\Phi \nabla^{\beta}\Phi - 2 J_9 \nabla_{\alpha}\Phi \nabla^{\alpha}\Phi \nabla_{\beta}\Phi \nabla^{\beta}\Phi\nonumber \\ 
&  + (-2 J_7 + 2 J_9) \nabla^{\alpha}\Phi \nabla_{\beta}\nabla_{\alpha}\Phi \nabla^{\beta}\Phi + J_{14} H_{\alpha}{}^{\gamma \delta} H_{\beta \gamma \delta} \nabla^{\beta}\nabla^{\alpha}\Phi + J_7 \nabla_{\beta}\nabla_{\alpha}\Phi \nabla^{\beta}\nabla^{\alpha}\Phi  \nonumber \\ 
& + J_{12} H^{\alpha \beta \gamma} \nabla_{\gamma}\nabla_{\delta}H_{\alpha \beta}{}^{\delta} + (-2 J_{11}+ J_{14}) H^{\beta \gamma \delta} \nabla^{\alpha}\Phi \nabla_{\delta}H_{\alpha \beta \gamma} + J_{12} \nabla_{\alpha}H^{\alpha \beta \gamma} \nabla_{\delta}H_{\beta \gamma}{}^{\delta}  \nonumber \\ 
& + (-2 J_{12} + J_{14}) H_{\alpha}{}^{\beta \gamma} \nabla^{\alpha}\Phi \nabla_{\delta}H_{\beta \gamma}{}^{\delta}+ J_{11} H^{\alpha \beta \gamma} \nabla_{\delta}\nabla_{\gamma}H_{\alpha \beta}{}^{\delta} + J_{10} H^{\alpha \beta \gamma} \nabla_{\delta}\nabla^{\delta}H_{\alpha \beta \gamma} \label{eq.04301916} \\ 
&  + J_{11} \nabla_{\gamma}H_{\alpha \beta \delta} \nabla^{\delta}H^{\alpha \beta \gamma}+ J_{10} \nabla_{\delta}H_{\alpha \beta \gamma} \nabla^{\delta}H^{\alpha \beta \gamma}+ J_{13} H_{\beta \gamma \delta} H^{\beta \gamma \delta} \nabla_{\alpha}\nabla^{\alpha}\Phi+ J_5 \nabla_{\beta}\nabla_{\alpha}\nabla^{\beta}\nabla^{\alpha}\Phi  \nn
\end{align}
One is free to add $\mathcal{J}_1$ to $\mathcal{L}_1$ and choose the parameters  $J_1,\cdots, J_{14}$ to reduce the couplings in  \eqref{eq.04301918}. 

The couplings in $\mathcal{J}_1+\mathcal{L}_1$, however, are in a  fixed field variables. One is free to change the field variables as 
\begin{eqnarray}
g_{\mu\nu}&\rightarrow &g_{\mu\nu}+\alpha' \delta g^{(1)}_{\mu\nu}\nn\\
B_{\mu\nu}&\rightarrow &B_{\mu\nu}+ \alpha'\delta B^{(1)}_{\mu\nu}\nn\\
\Phi &\rightarrow &\Phi+ \alpha'\delta\Phi^{(1)}\label{gbp}
\end{eqnarray}
where the tensors $\delta g^{(1)}_{\mu\nu}$, $\delta B^{(1)}_{\mu\nu}$ and $\delta\Phi^{(1)}$ are all possible covariant and gauge invariant terms at two-derivative level, \ie,
\begin{align}
 \delta g^{(1)}_{\mu\nu}=& a_1^{\text{}} R_{\mu \nu}+ a_2^{\text{}} H_{\mu}{}^{\alpha \beta} H_{\nu \alpha \beta}+ a_3^{\text{}} \nabla_{\nu}\nabla_{\mu}\Phi+ a_4^{\text{}} \nabla_{\mu}\Phi \nabla_{\nu}\Phi + g_{\mu \nu}\Big( a_5^{\text{}} R +a_6^{\text{}} H_{\alpha \beta \gamma} H^{\alpha \beta \gamma}\nn\\
& + a_7^{\text{}}  \nabla_{\alpha}\nabla^{\alpha}\Phi +  a_8 \nabla_{\alpha}\Phi \nabla^{\alpha}\Phi \Big) \nn\\
 \delta B^{(1)}_{\mu\nu}=& a_9^{\text{}} \nabla_{\alpha}H_{\mu \nu}{}^{\alpha} + a_{10}^{\text{}} H_{\mu \nu \alpha} \nabla^{\alpha}\Phi \nn\\
 \delta\Phi^{(1)}=&a_{11}^{\text{}} R +a_{12}^{\text{}} H_{\alpha \beta \gamma} H^{\alpha \beta \gamma} + a_{13}^{\text{}}  \nabla_{\alpha}\nabla^{\alpha}\Phi +  a_{14} \nabla_{\alpha}\Phi \nabla^{\alpha}\Phi \label{eq.12}
\end{align}
The coefficients $a_1,\cdots, a_{14}$ are arbitrary parameters. When the field variables in $\sqrt{-g}e^{-2\Phi}(\mathcal{J}_1+\mathcal{L}_1)$  are changed according to the above  field redefinitions, they produce some couplings at order $\alpha'^2$ in which we are not interested in this section. However, when the field variables in $S_0$  are changed,  up to some total derivative terms, the following   couplings  at order $\alpha'$ are produced:
\begin{align}
\delta S_0=&\frac{\delta S_0}{\delta g_{\alpha\beta}}\delta g^{(1)}_{\alpha\beta}+\frac{\delta S_0}{\delta B_{\alpha\beta}}\delta B^{(1)}_{\alpha\beta}+\frac{\delta S_0}{\delta \Phi}\delta \Phi^{(1)}\equiv \int d^Dx\sqrt{-g}e^{-2\Phi}\mathcal{K}_1\nn\\
=& \int d^Dx\sqrt{-g}e^{-2\Phi}\Big[(\tfrac{1}{2} \nabla_{\gamma}H^{\alpha \beta \gamma} -  H^{\alpha \beta}{}_{\gamma} \nabla^{\gamma}\Phi)\delta B^{(1)}_{\alpha\beta} -(  R^{\alpha \beta}-\tfrac{1}{4} H^{\alpha \gamma \delta} H^{\beta}{}_{\gamma \delta}+ 2 \nabla^{\beta}\nabla^{\alpha}\Phi)\delta g^{(1)}_{\alpha\beta}
\nn\\
&-2( R -\tfrac{1}{12} H_{\alpha \beta \gamma} H^{\alpha \beta \gamma} + 4 \nabla_{\alpha}\nabla^{\alpha}\Phi -4 \nabla_{\alpha}\Phi \nabla^{\alpha}\Phi)(\delta\Phi^{(1)}-\frac{1}{4}\delta g^{(1)\mu}{}_\mu) \Big]\label{eq.13}
\end{align}
  Replacing \eqref{eq.12} into \eqref{eq.13}, one finds
\begin{align}
\mathcal{K}_1=& 
  -  a_1 R_{\alpha \beta} R^{\alpha \beta}   + \tfrac{1}{2} \bigl(a_1 - 4 a_{11} + a_5 (-2 + D)\bigr) R^2 
+\tfrac{1}{4} a_2 H_{\alpha \beta}{}^{\delta} H^{\alpha \beta \gamma} H_{\gamma}{}^{\epsilon \zeta} H_{\delta \epsilon \zeta}\nn\\
& + \tfrac{1}{24} \bigl(4 a_{12} -  a_2 -  a_6 (-6 + D)\bigr) H_{\alpha \beta \gamma} H^{\alpha \beta \gamma} H_{\delta \epsilon \zeta} H^{\delta \epsilon \zeta} 
 + \tfrac{1}{4} (a_1  - 4 a_2) H_{\alpha}{}^{\gamma \delta} H_{\beta \gamma \delta} R^{\alpha \beta}\nn\\
&  + \tfrac{1}{24} (- a_1 + 4 a_{11} - 48 a_{12} + 12 a_2 + 6 a_5 - 24 a_6 -  a_5 D + 12 a_6 D) H_{\alpha \beta \gamma} H^{\alpha \beta \gamma} R\nn\\
&  + \tfrac{1}{24} (-192 a_{12} + 4 a_{13} + 48 a_2 -  a_3 - 48 a_6 + 6 a_7 + 48 a_6 D -  a_7 D) H_{\beta \gamma \delta} H^{\beta \gamma \delta} \nabla_{\alpha}\nabla^{\alpha}\Phi \nn\\
& + \tfrac{1}{2} (4 a_1 - 16 a_{11} - 4 a_{13} + a_3 - 4 a_5 - 2 a_7 + 4 a_5 D + a_7 D) R \nabla_{\alpha}\nabla^{\alpha}\Phi \nn\\
& + \tfrac{1}{24} (192 a_{12} + 4 a_{14} - 48 a_2 -  a_4 + 6 a_8 - 48 a_6 D -  a_8 D) H_{\beta \gamma \delta} H^{\beta \gamma \delta} \nabla_{\alpha}\Phi \nabla^{\alpha}\Phi \nn\\
& + \tfrac{1}{2} (-4 a_1 + 16 a_{11} - 4 a_{14} + a_4 - 2 a_8 - 4 a_5 D + a_8 D) R \nabla_{\alpha}\Phi \nabla^{\alpha}\Phi 
 + (-2 a_1 -  a_3) R^{\alpha \beta} \nabla_{\beta}\nabla_{\alpha}\Phi \nn\\
& + 2 \bigl(-4 a_{13} + a_3 + a_7 (-1 + D)\bigr) \nabla_{\alpha}\nabla^{\alpha}\Phi \nabla_{\beta}\nabla^{\beta}\Phi  + \tfrac{1}{4} (-4 a_{10} + a_4) H_{\alpha}{}^{\gamma \delta} H_{\beta \gamma \delta} \nabla^{\alpha}\Phi \nabla^{\beta}\Phi\nn\\
& + 2 (4 a_{13} - 4 a_{14} -  a_3 + a_4 -  a_8 -  a_7 D  + a_8 D) \nabla_{\alpha}\Phi \nabla^{\alpha}\Phi \nabla_{\beta}\nabla^{\beta}\Phi 
  -  a_4 R_{\alpha \beta} \nabla^{\alpha}\Phi \nabla^{\beta}\Phi \nn\\
  &+ \bigl(8 a_{14} - 2 (a_4  + a_8 D)\bigr) \nabla_{\alpha}\Phi \nabla^{\alpha}\Phi \nabla_{\beta}\Phi \nabla^{\beta}\Phi 
  - 2 a_4 \nabla^{\alpha}\Phi \nabla_{\beta}\nabla_{\alpha}\Phi \nabla^{\beta}\Phi \nn\\
 & + \tfrac{1}{4} (-8 a_2 + a_3) H_{\alpha}{}^{\gamma \delta} H_{\beta \gamma \delta} \nabla^{\beta}\nabla^{\alpha}\Phi  
 - 2 a_3 \nabla_{\beta}\nabla_{\alpha}\Phi \nabla^{\beta}\nabla^{\alpha}\Phi \nn\\
&+ \tfrac{1}{2} a_9 \nabla_{\alpha}H^{\alpha \beta \gamma} \nabla_{\delta}H_{\beta \gamma}{}^{\delta} 
+ \tfrac{1}{2} (a_{10} - 2 a_9) H_{\alpha}{}^{\beta \gamma} \nabla^{\alpha}\Phi \nabla_{\delta}H_{\beta \gamma}{}^{\delta}\label{eq.04301917}
\end{align}
where $D$ is the  number of spacetime dimensions.

Not all arbitrary parameters $a_1,\cdots,a_{14}$  produce non-zero $\mathcal{K}_1$. For some relations between the parameters, one finds the field redefinition \eqref{gbp} is the general  coordinate transformation which obviously leaves   $\mathcal{K}_1$  invariant up to some boundary term. In fact under the  coordinate transformation  $x^\mu\to x^\mu+\varepsilon^\mu=x^\mu+a\nabla^\mu\Phi$, one has the following transformations for fields:
\begin{align}
\delta g_{\mu\nu}&=\nabla_\mu\varepsilon_\nu+\nabla_\nu\varepsilon_\mu=2a\nabla_{\mu}\nabla_\nu\Phi \nn\\
\delta \Phi&=\varepsilon^\mu\nabla_\mu\Phi=a \nabla^\mu\Phi\nabla_\mu\Phi \label{eq.01121815}\\
\delta B_{\mu\nu}&=\varepsilon^\gamma H_{\gamma\mu\nu}=aH_{\alpha\mu\nu}\nabla^\alpha\Phi\nn
\end{align}
Hence, if $a_3=2a_{10}$ and $a_{14}=a_{10}$, the corresponding field redefinitions in \eqref{eq.12} is a coordinate transformation which leaves $\mathcal{K}_1$ invariant, up to some boundary term.
For some other relations between the parameters,  $\mathcal{K}_1$ may still be  invariant, however, the corresponding transformation is not the coordinate transformation. If one  removes the parameters that leave $\mathcal{K}_1$ invariant,  then the remaining parameters   all would be  fixed after using the field redefinitions.  On the other hand, if one  keeps    all parameters  $a_1,\cdots,a_{14}$, then   some of the parameters remain  arbitrary after using the field redefinitions.  We use this latter method and work with all parameters  $a_1,\cdots,a_{14}$.  

One is free to add $\mathcal{K}_1$ to $\mathcal{L}_1+\mathcal{J}_1$ and choose the parameters  $J_1,\cdots, J_{14}$ and $a_1,\cdots, a_{14}$ to reduce the couplings in  \eqref{eq.04301918}. The Bianchi identities and the commutation relation of the covariant derivatives, however, are not  yet used in  $\mathcal{L}_1+\mathcal{J}_1+\mathcal{K}_1$. They are
\begin{align}
& R_{\alpha[\beta\gamma\delta]}=0,\nn\\
& \nabla_{[\mu}R_{\alpha\beta]\gamma\delta}=0,\label{eq.01121726}\\
&\nabla_{[\mu}H_{\alpha\beta\gamma]}=0,\nn\\
&[\nabla,\nabla]\mathcal{O}=R\mathcal{O}\nn
\end{align}
They can  further reduce  the independent couplings in  \eqref{eq.04301918}.  The above identities may be contracted with tensors $R,H,\nabla\Phi$ and their derivatives with arbitrary parameters and then add them to $\mathcal{L}_1+\mathcal{J}_1+\mathcal{K}_1$.  By manipulating the arbitrary parameters, one may find the independent couplings in   \eqref{eq.04301918}. Instead of imposing  the identities \eqref{eq.01121726} with arbitrary parameters, we  use the locally inertial frame in which the above identities are almost  automatically satisfied. 

In the locally inertial frame, the metric $g_{\alpha\beta}$ takes its canonical form and its first derivatives are all vanish, \ie,
\[
g_{\alpha\beta}=\eta_{\alpha\beta},\qquad \pa_\mu g_{\alpha\beta}=0
\]
The second and higher derivatives of metric, however, are non-zero. In this coordinate,   by rewriting the covariant derivative in terms of partial derivatives, one finds, except the Bianchi identity $dH=0$, all other identities  in \eqref{eq.01121726} are satisfied. To satisfy  the Bianchi identity $dH=0$ as well, in the couplings which involve derivatives of $H$, we   rewrite    $H$ in terms of   $B$-field, \ie, $H_{\mu\nu\alpha}=\pa_\mu B_{\nu\alpha}+\pa_\alpha B_{\mu\nu}+\pa_\nu B_{\alpha\mu}$. When writing the couplings  $\mathcal{L}_1+\mathcal{J}_1+\mathcal{K}_1$ in this local frame,  then all resulting terms in  $\mathcal{L}_1+\mathcal{J}_1+\mathcal{K}_1$ become  independent. Using the arbitrary parameters   $J_1,\cdots, J_{14}$ and $a_1,\cdots, a_{14}$, one may find the    couplings in many different schemes. 

To clarify this point,   one may write the final form of the couplings as $\mathcal{L}_1'$ which is the same as the couplings \eqref{eq.04301918} with different parameters $B_1',\cdots, B_{41}'$. Then writing 
\begin{align}
\mathcal{L}_1'-\mathcal{L}_1=\mathcal{J}_1+\mathcal{K}_1\label{LL}
\end{align}
in the local frame, one finds some relations between the arbitrary parameters  $J_1,\cdots, J_{14}$, $a_1,\cdots, a_{14}$  and  $\delta B_1,\cdots, \delta B_{41}$ where $\delta B_i=B_i'-B_i$. These are very lengthy expressions that we do not write them here explicitly. There are two parameters $a_{10},a_8$ in these relations which remain unfixed.  

 The equation \eqref{LL} produces the following 8 relations between only $\delta B_1,\cdots, \delta B_{41}$ as well:
\begin{align}
&\quad \delta B_{22} =0\nn\\ 
&\quad 2 \delta B_1  + \delta B_2 =0\label{dB}\\
&\quad  16 \delta B_4  - 8 \delta B_9-4 \delta B_{10}  + 4 \delta B_{18}  + 2 \delta B_{20}  + \delta B_{21}  =0\nn\\
&\quad 6 \delta B_{25}  + 2 \delta B_{26}  - 6 \delta B_{28}  - 2 \delta B_{30}  - 2 \delta B_{31}  - \delta B_{32}  = 0\nn\\
&\quad \delta B_3+ 16 \delta B_{23}  + 12 \delta B_{25}  + 4 \delta B_{26}  - 12 \delta B_{28}    - 4 \delta B_{30}  + 4 \delta B_{33} =0\nn\\
&\quad 2 \delta B_{14}  + \delta B_{16}  -  \delta B_{17}  + 16 \delta B_{29}  - 4 \delta B_3  + 8 \delta B_{37}  + 4 \delta B_{41}  + 2 \delta B_7  + \delta B_8 =0\nn\\
&\quad 10 \delta B_{10}  - 4 \delta B_{11}  - 8 \delta B_{12}  + 8 \delta B_{13}  + 4 \delta B_{14}  + 4 \delta B_{15}  + 2 \delta B_{16}  - 8 \delta B_{18}  + \delta B_{19}  - 2 \delta B_{20} \nn\\
&\quad - 96 \delta B_{34}  + 48 \delta B_{38}  - 80 \delta B_4  + 24 \delta B_{40}  - 8 \delta B_5  - 16 \delta B_6  - 2 \delta B_8  + 28 \delta B_9 =0\nn\\
&\quad 5 \delta B_{11}  + 10 \delta B_{12}  - 4 \delta B_{13}  - 4 \delta B_{14}  - 2 \delta B_{15}  - 2 \delta B_{16}  + 2 \delta B_{17}  + 2 \delta B_{18}  + 288 \delta B_{24}  + 24 \delta B_{25}\nn\\
&\quad   + 8 \delta B_{26}  + 12 \delta B_3  + 8 \delta B_{33}  + 120 \delta B_{34}  + 12 \delta B_{35}  + 4 \delta B_{36}  - 24 \delta B_{38}  - 4 \delta B_{39}  + 50 \delta B_4  + 10 \delta B_5 \nn\\
&\quad + 20 \delta B_6  - 5 \delta B_7  - 10 \delta B_9 =0\nn
\end{align}
The first relation, $\delta B_{22}=0$, indicates that the parameter $B_{22}'=B_{22}$ is not changed by the field redefinition, by adding total derivative term  or by using the Bianchi identities. It is an unambiguous parameter. All other relations indicate that the other parameters are ambiguous parameters because they are changed by the field redefinition, by  total derivative term  or by  the Bianchi identities. 

To find the minimum number of couplings in $\mathcal{L}_1'$, one may choose 33 parameters  in  $\mathcal{L}_1'$ to be zero. These parameters, however, should change the 8 equations in \eqref{dB} to  8 equations  $\delta B_i=f_i(B_1,\cdots,B_{41})$ where  $i$ is 8 numbers  among $1,\cdots, 41$ depending on the scheme that one uses for the terms in  $\mathcal{L}_1'$ to be zero. It is obvious that one of them is  $i=22$ for which   $f_{22}=0$.  If one chooses $B_2'=0$, then the second equation above indicates that $\delta B_1=B_2/2$. Alternatively, if one chooses $B_1'=0$, then the second equation becomes  $\delta B_2=2B_1$. Similarly for all other equations which have many different schemes.  In any scheme, the non-zero parameters in $\mathcal{L}_1'$ are $B_i'=B_i+f_i(B_1,\cdots,B_{41})$.

We choose a set of  zero  parameters in $\mathcal{L}_1'$ to be those that their corresponding couplings have terms with more than two derivatives or have $R$,$R_{\mu\nu}$, $\nabla_{\mu}H^{\mu\alpha\beta}$, $\nabla_\mu\nabla^\mu\Phi$. There are, however, 24 such parameters. One may  set to zero the other 9 parameters in $\mathcal{L}_1'$ such  that the remaining non-zero parameters becomes the one considered in \cite{Metsaev:1987zx}, \ie,  $B_1',B_{21}', B_{22}', B_{23}',B_{24}', B_{31}', B_{40}', B_{41}'$. 

However, we choose two   other  schemes. In one scheme, we write the couplings as the following:
\begin{align}
 \mathcal{L}'_1= \mathcal{L}^1_1+ \mathcal{L}^2_1\label{L1112}
\end{align}
where $\mathcal{L}^1_1$ includes the minimum number of couplings which do not include the dilaton, \ie,
\begin{eqnarray}
\mathcal{L}^1_1&= &  B_1 R_{\alpha \gamma \beta \delta} R^{\alpha \beta \gamma \delta}
  +B_2 H_{\alpha}{}^{\delta \epsilon} H^{\alpha \beta \gamma} H_{\beta \delta}{}^{\zeta} H_{\gamma \epsilon \zeta} \nn\\
  &&
  + B_3 H_{\alpha \beta}{}^{\delta} H^{\alpha \beta \gamma} H_{\gamma}{}^{\epsilon \zeta} H_{\delta \epsilon \zeta} 
 + B_4 H_{\alpha}{}^{\delta \epsilon} H^{\alpha \beta \gamma} R_{\beta \delta \gamma \epsilon}\label{L11}
 \end{eqnarray}
 and  $\mathcal{L}^2_1$ includes the other couplings  which   all include non-tivially the dilaton, \ie,
 \begin{eqnarray}
  \mathcal{L}^2_1&= & B_5 H_{\beta \gamma \delta} H^{\beta \gamma \delta} \nabla_{\alpha}\Phi \nabla^{\alpha}\Phi
 + B_6 H_{\alpha}{}^{\gamma \delta} H_{\beta \gamma \delta} \nabla^{\alpha}\Phi \nabla^{\beta}\Phi \nn\\&&
 + B_7 H_{\alpha}{}^{\gamma \delta} H_{\beta \gamma \delta} \nabla^{\beta}\nabla^{\alpha}\Phi
 + B_8 \nabla_{\alpha}\Phi \nabla^{\alpha}\Phi \nabla_{\beta}\Phi \nabla^{\beta}\Phi \label{L12}
 \end{eqnarray}
 where we have also dropped the prime on the coefficients and relabelled then from 1 to 8. The reason for  the couplings in \eqref{L11} to be   the minimum number of couplings which do not include the dilaton, is that when one sets the dilaton to be constant, there are only 4  relations between $\delta B$s. 
 
  In the other scheme, we write the couplings as the following:
\begin{align}
 \mathcal{L}'_1= \mathcal{L}^1_1+ \mathcal{L}^3_1\label{L1113}
\end{align}
where $\mathcal{L}^1_1$ is the same as in \eqref{L11}, and $\mathcal{L}^3_1$ includes   4 couplings other than those appear  in  $\mathcal{L}^1_1$ in which the dilaton do not appear, \ie, 
 \begin{align}
 \mathcal{L}^3_1= &  B_5 R^2 
  + B_6 H_{\alpha \beta \gamma} H^{\alpha \beta \gamma} R 
 + B_7 \nabla_{\alpha}H^{\alpha \beta \gamma} \nabla_{\delta}H_{\beta \gamma}{}^{\delta}
 + B_8 H_{\alpha \beta \gamma} H^{\alpha \beta \gamma} H_{\delta \epsilon \zeta} H^{\delta \epsilon \zeta}
 \end{align}
 The 8 parameters in \eqref{L1112} or \eqref{L1113} have been fixed by comparison with the three- and four-point string amplitudes \cite{Metsaev:1987zx}. They have been also fixed, up to an overall factor,  by the T-duality constraint \cite{Garousi:2019wgz}. Only the  parameters in $\mathcal{L}_1^1$ are non-zero!  

The parameters in the field redefinitions  $\delta g^{(1)}_{\alpha\beta},\delta B^{(1)}_{\alpha\beta}$ and $\delta\Phi^{(1)}$ that change the action \eqref{eq.04301918} to \eqref{L1112} or \eqref{L1113}, are functions of $\delta B_1,\cdots, \delta B_{41}$ and $a_{10},a_8$. In the above schemes, $\delta B_i=f_i(B_1,\cdots, B_{41})$ where $i $ is 8 numbers among $1,\cdots, 41$, and all others are $\delta B_j=-B_j$. The parameter $a_{10}$ produces coordinate transformations in which we are not interested, and parameters $a_8$ produces the transformation that leaves $\mathcal{K}_1$ invariant. We will discuss more about this term in the next section.

\section{Minimal  couplings at order  $\alpha'^2$}\label{sec.3}

In this section we extend the calculations in the previous section to the order $\alpha'^2$.  We begin by writing all  possible covariant and gauge invariant couplings 
  at six-derivative order, \ie,
\begin{align}
\mathcal{L}_2=&C_1 R_{\alpha}{}^{\epsilon}{}_{\gamma}{}^{\zeta} R^{\alpha \beta \gamma \delta} R_{\beta \zeta \delta \epsilon} 
+ C_2 R_{\alpha \beta}{}^{\epsilon \zeta} R^{\alpha \beta \gamma \delta} R_{\gamma \epsilon \delta \zeta} 
+ C_3  H_{\alpha}{}^{\gamma \delta} H_{\gamma}{}^{\epsilon \zeta} R_{\beta \epsilon \delta \zeta} \nabla^{\beta}\nabla^{\alpha}\Phi  +\cdots  
 \label{eq.final0}
\end{align}
There
  are 705 such couplings, however, they are not independent. To remove the total derivative terms from the above couplings, we consider the most general total derivative terms at order $\alpha'^2$ which has the following structure:
\begin{align}
\alpha'^2\int d^Dx \sqrt{-g}e^{-2\Phi} \mathcal{J}_2=\alpha'^2\int d^Dx\sqrt{-g} \nabla_\alpha (e^{-2\Phi}J_2^\alpha) \label{eq.043019}
\end{align}
where the vector $J_2^\alpha$ is   all possible covariant and gauge invariant  terms at five-derivative level, \ie,
\begin{align}
J_2^\alpha=&J_1\nabla^\alpha\Phi R^2+J_2H^{\alpha\mu\nu}H_{\beta\mu\nu}\nabla^\beta\Phi R+\cdots
\end{align}
There are 315 such terms with arbitrary parameters. The corresponding $\mathcal{J}_2$ has 641 terms.

Now we consider the field redefinitions at order $\alpha'^2$. Under the field redefinitions
\begin{eqnarray}
g_{\mu\nu}&\rightarrow &g_{\mu\nu}+\alpha' \delta g^{(1)}_{\mu\nu}+\alpha'^2 \delta g^{(2)}_{\mu\nu}\nn\\
B_{\mu\nu}&\rightarrow &B_{\mu\nu}+ \alpha'\delta B^{(1)}_{\mu\nu}+ \alpha'^2\delta B^{(2)}_{\mu\nu}\nn\\
\Phi&\rightarrow &\Phi+ \alpha'\delta\Phi^{(1)}+ \alpha'^2\delta\Phi^{(2)}\label{gbp2}
\end{eqnarray}
where the deformations at orders $\alpha'$ and $\alpha'^2$ are arbitrary, the actions $S_0$ and $S_1$ produces the following contributions at order $\alpha'^2$, up to some total derivative terms: 
\begin{eqnarray}
\delta S_0+\delta S_1&=&\frac{\delta S_0}{\delta g_{\alpha\beta}}\delta g^{(2)}_{\alpha\beta}+\frac{\delta S_0}{\delta B_{\alpha\beta}}\delta B^{(2)}_{\alpha\beta}+\frac{\delta S_0}{\delta \Phi}\delta \Phi^{(2)}
+\frac{\delta S_1}{\delta g_{\alpha\beta}}\delta g^{(1)}_{\alpha\beta}+\frac{\delta S_1}{\delta B_{\alpha\beta}}\delta B^{(1)}_{\alpha\beta}+\frac{\delta S_1}{\delta \Phi}\delta \Phi^{(1)}\nn\\
&&+S_0(\delta g^{(1)},\delta g^{(1)})+S_0 (\delta g^{(1)},\delta B^{(1)})+S_0 (\delta g^{(1)},\delta\Phi^{(1)})\nn\\
&&+S_0(\delta B^{(1)},\delta B^{(1)})+S_0 (\delta B^{(1)},\delta \Phi^{(1)})+S_0 (\delta \Phi^{(1)},\delta\Phi^{(1)})\label{eq.field.1}
\end{eqnarray}
where $S_0(\delta g^{(1)},\delta g^{(1)})$     which includes  $\delta g^{(1)}\delta g^{(1)}$-terms, is resulted from replacing the  transformation $g\to g+\alpha'\delta g^{(1)}, B\to B+\alpha'\delta B^{(1)}$ and $\Phi\to \Phi+\alpha'\delta \Phi^{(1)}$  into  $S_0$. Similarly for all other terms in the second and the third line above. Up to some total derivative terms, one may write $S_0(\delta g^{(1)},\delta g^{(1)})=(\cdots)\delta g^{(1)}$. Similarly for other terms in \eqref{eq.field.1}. As a result, one may rewrite  the above equation as
\begin{align}
\delta S_0+\delta S_1=&\frac{\delta S_0}{\delta g_{\alpha\beta}}\delta g^{(2)}_{\alpha\beta}+\frac{\delta S_0}{\delta B_{\alpha\beta}}\delta B^{(2)}_{\alpha\beta}+\frac{\delta S_0}{\delta \Phi}\delta \Phi^{(2)}\nn\\
&+(\frac{\delta S_1}{\delta g_{\alpha\beta}}+\cdots)\delta g^{(1)}_{\alpha\beta}+(\frac{\delta S_1}{\delta B_{\alpha\beta}}+\cdots)\delta B^{(1)}_{\alpha\beta}+(\frac{\delta S_1}{\delta \Phi}+\cdots)\delta \Phi^{(1)} \label{eq.field.2}
\end{align}
However, in the previous section, we have  adjusted $\delta g^{(1)}_{\alpha\beta},\delta B^{(1)}_{\alpha\beta}$ and $\delta\Phi^{(1)}$ so as to obtain the   action    $S_{1}$   with fixed parameters. All parameters in \eqref{eq.12} are fixed except the  two parameters $a_{10},a_8$. The parameter $a_{10}$ which produces the coordinate transformation should not be included in the field redefinitions,   and the parameter $a_8$ which   leave $\mathcal{K}_1$ invariant but is not corresponding to the coordinate transformation may be included in the field redefinition. We call the corresponding field deformations   $\delta \hat{g}^{(1)}_{\alpha\beta},\delta \hat{B}^{(1)}_{\alpha\beta}$ and $\delta\hat{\Phi}^{(1)}$. In fact the equation $\mathcal{K}_1=0$ has the following solution: 
\begin{align}
&a_1=a_2=a_4=a_9=a_{10}=0\nn\\
&-4a_5=48a_6=-a_7=-\frac{16}{D-2}a_{11}=\frac{192}{D-6}a_{12}=-\frac{4}{D-1}a_{13}=\frac{4}{D}a_{14}=a_8\nn
\end{align}
The corresponding field deformations are 
\begin{align}
  \delta \hat{g}^{(1)}_{\mu\nu}=& -a_8 g_{\mu \nu}\Big(\frac{1}{4} R -\frac{1}{48} H_{\alpha \beta \gamma} H^{\alpha \beta \gamma}+ \nabla_{\alpha}\nabla^{\alpha}\Phi -  \nabla_{\alpha}\Phi \nabla^{\alpha}\Phi \Big) \nn\\
  \delta \hat{B}^{(1)}_{\mu\nu}=&0\label{eq.14021053}\\
 \delta\hat{\Phi}^{(1)}=&-\frac{a_8}{4}\Big(\frac{D-2}{4} R -\frac{D-6}{48} H_{\alpha \beta \gamma} H^{\alpha \beta \gamma} +(D-1)  \nabla_{\alpha}\nabla^{\alpha}\Phi - D \nabla_{\alpha}\Phi \nabla^{\alpha}\Phi \Big)\nn
\end{align}
Since we have already fixed the field redefinitions at order $\alpha'$ to choose the schemes  \eqref{L1112} or \eqref{L1113}, one should consider only the above residual deformations in \eqref{eq.field.2}.    

Up to some total derivative terms, \eqref{eq.14021053}  can be written as 
\begin{align}
\int d^Dx\sqrt{-g}e^{-2\Phi} \delta \hat{g}^{(1)}_{\mu\nu}=&\frac{1}{8}a_8g_{\mu\nu}\frac{\delta S_0}{\delta \Phi}\nn\\
 \delta \hat{B}^{(1)}_{\mu\nu}=&0\label{PgB1}\\
  \int d^Dx\sqrt{-g}e^{-2\Phi}\delta\hat{\Phi}^{(1)}=&\frac{D}{8}a_8\frac{\delta S_0}{\delta \Phi}-\frac{1}{8}a_8g_{\mu\nu}\frac{\delta S_0}{\delta g_{\mu\nu}}\nn
\end{align}
Replacing it into  \eqref{eq.field.2}, one can rewrite   \eqref{eq.field.2} as 
\begin{align}
&\delta S_0+\delta S_1=\frac{\delta S_0}{\delta g_{\alpha\beta}}\delta g'^{(2)}_{\alpha\beta}+\frac{\delta S_0}{\delta B_{\alpha\beta}}\delta B^{(2)}_{\alpha\beta}+\frac{\delta S_0}{\delta \Phi}\delta \Phi'^{(2)}\equiv \int d^Dx\sqrt{-g}e^{-2\Phi}\mathcal{K}_2\nn\\
=& \int d^Dx\sqrt{-g}e^{-2\Phi}\Big[(\tfrac{1}{2} \nabla_{\gamma}H^{\alpha \beta \gamma} -  H^{\alpha \beta}{}_{\gamma} \nabla^{\gamma}\Phi)\delta B^{(2)}_{\alpha\beta} -(  R^{\alpha \beta}-\tfrac{1}{4} H^{\alpha \gamma \delta} H^{\beta}{}_{\gamma \delta}+ 2 \nabla^{\beta}\nabla^{\alpha}\Phi)\delta g'^{(2)}_{\alpha\beta}
\nn\\
&-2( R -\tfrac{1}{12} H_{\alpha \beta \gamma} H^{\alpha \beta \gamma} + 4 \nabla_{\alpha}\nabla^{\alpha}\Phi -4 \nabla_{\alpha}\Phi \nabla^{\alpha}\Phi)(\delta\Phi'^{(2)}-\frac{1}{4}\delta g'^{(2)\mu}{}_\mu) \Big]  \label{eq.field3}
\end{align}
where the deformations $\delta g'^{(2)}_{\alpha\beta},\delta \Phi'^{(2)}$ and   $\delta g^{(2)}_{\alpha\beta},\delta \Phi^{(2)}$ have different parameters, however, since we have not yet fixed the parameters in  $\delta g^{(2)}_{\alpha\beta},\delta \Phi^{(2)}$, we consider the field redefinitions 
\begin{eqnarray}
g_{\mu\nu}&\rightarrow &g_{\mu\nu}+\alpha' \delta g^{(1)}_{\mu\nu}+\alpha'^2 \delta g'^{(2)}_{\mu\nu}\nn\\
B_{\mu\nu}&\rightarrow &B_{\mu\nu}+ \alpha'\delta B^{(1)}_{\mu\nu}+ \alpha'^2\delta B^{(2)}_{\mu\nu}\nn\\
\Phi&\rightarrow &\Phi+ \alpha'\delta\Phi^{(1)}+ \alpha'^2\delta\Phi'^{(2)}\label{gbp3}
\end{eqnarray}
in which the deformations at order $\alpha'$ are all already fixed to find the action  \eqref{L1112} or \eqref{L1113}, and the deformations at order $\alpha'^2$ are yet arbitrary.

The most general deformations at order $\alpha'^2$ are:
\begin{align}
\delta\Phi'^{(2)}=&b_1^{\text{}} H_{\alpha}{}^{\delta \epsilon} H^{\alpha \beta \gamma} H_{\beta \delta}{}^{\zeta} H_{\gamma \epsilon \zeta} + b_2^{\text{}} H_{\alpha \beta}{}^{\delta} H^{\alpha \beta \gamma} H_{\gamma}{}^{\epsilon \zeta} H_{\delta \epsilon \zeta} + \cdots \nn\\
\delta B^{(2)}_{\mu\nu}=&c_{1}^{\text{}} R_{\mu \nu \beta \gamma} \nabla_{\alpha}H^{\alpha \beta \gamma} + c_2^{\text{}} H^{\beta \gamma \delta} H_{\mu \nu}{}^{\alpha} \nabla_{\alpha}H_{\beta \gamma \delta} +\cdots \nn\\ 
\delta g'^{(2)}_{\mu\nu}=&d_1^{\text{}} H_{\gamma \delta \epsilon} H^{\gamma \delta \epsilon} H_{\mu}{}^{\alpha \beta} H_{\nu \alpha \beta} + d_2^{\text{}} H_{\beta}{}^{\delta \epsilon} H_{\gamma \delta \epsilon} H_{\mu}{}^{\alpha \beta} H_{\nu \alpha}{}^{\gamma} + \cdots \label{eq.g.1}
\end{align}
where $b_1,\cdots, b_{41}$, $ c_1,\cdots c_{81}$ and $d_1,\cdots, d_{121}$ are arbitrary parameters. 

To find the independent couplings, we write the final form of the couplings as $\mathcal{L}_2'$ which is the same as the couplings \eqref{eq.final0} with different parameters $C_1',\cdots, C_{705}'$. Then writing 
\begin{align}
\mathcal{L}_2'-\mathcal{L}_2=\mathcal{J}_2+\mathcal{K}_2\label{LL2}
\end{align}
in the local frame, one finds some relations between the arbitrary parameters of $\mathcal{L}_2,\,\mathcal{K}_2$  and  $\delta C_1,\cdots, \delta C_{705}$  in which we are not interested, and also 60 relations between only  $\delta C_1,\cdots, \delta C_{705}$ in which we are interested. Note that these relations are independent of the form of the fixed action at order $\alpha'$, whether it is minimal action or not (see \cite{Bento1990} for the case that $B$-field is zero).   

To find the minimum number of couplings in $\mathcal{L}_2'$, one may choose 645 parameters  in  $\mathcal{L}_2'$ to be zero. These parameters, however, should change the 60 equations among  $\delta C_1,\cdots, \delta C_{705}$ to 60 equations  $\delta C_i=g_i(C_1,\cdots,C_{705})$ where  $i$ is 60 numbers  among $1,\cdots,705$ depending on the scheme that one uses for the terms in  $\mathcal{L}_2'$ to be zero.  As in the previous section we choose two schemes.  

In one scheme, we choose   a set of  zero  parameters in $\mathcal{L}_2'$ to be those that their corresponding couplings have terms with more than two derivatives or have $R$,$R_{\mu\nu}$, $\nabla_{\mu}H^{\mu\alpha\beta}$, $\nabla_\mu\nabla^\mu\Phi$. There  are 543 such couplings. We have found that it is consistent to set these parameters to zero, \ie, after setting  these parameters to zero, there are still 60 equations between the reaming $\delta C$s. There are still different schemes  for  choosing  the remaining  102 parameters in  $\mathcal{L}_2'$ to be zero. 
We choose the scheme in which   the minimum number of  couplings  in  $\mathcal{L}_2'$ to be: 
\begin{eqnarray}
\mathcal{L}_2'&=&\mathcal{L}_2^1+\mathcal{L}_2^2 \label{L2122}
\end{eqnarray}
where $\mathcal{L}_2^1$ has the minimum number of couplings in which the dilaton does not appear, \ie,
\begin{eqnarray}
\mathcal{L}_2^1&=& C_1R_{\alpha}{}^{\epsilon}{}_{\gamma}{}^{\zeta} R^{\alpha \beta \gamma \delta} R_{\beta \zeta \delta \epsilon} + C_2 R_{\alpha \beta}{}^{\epsilon \zeta} R^{\alpha \beta \gamma \delta} R_{\gamma \epsilon \delta \zeta}+C_3 H_{\alpha}{}^{\delta \epsilon} H^{\alpha \beta \gamma} H_{\beta \delta}{}^{\zeta} H_{\gamma}{}^{\iota \kappa} H_{\epsilon \iota}{}^{\mu} H_{\zeta \kappa \mu}\nn\\&& + C_4 H_{\alpha \beta}{}^{\delta} H^{\alpha \beta \gamma} H_{\gamma}{}^{\epsilon \zeta} H_{\delta}{}^{\iota \kappa} H_{\epsilon \zeta}{}^{\mu} H_{\iota \kappa \mu} + C_{5} H_{\alpha \beta}{}^{\delta} H^{\alpha \beta \gamma} H_{\gamma}{}^{\epsilon \zeta} H_{\delta \epsilon}{}^{\iota} H_{\zeta}{}^{\kappa \mu} H_{\iota \kappa \mu}\nn\\&& + C_{6} H_{\alpha}{}^{\delta \epsilon} H^{\alpha \beta \gamma} H_{\beta}{}^{\zeta \iota} H_{\delta \zeta}{}^{\kappa} R_{\gamma \epsilon \iota \kappa}  + C_{7} H_{\alpha}{}^{\delta \epsilon} H^{\alpha \beta \gamma} R_{\beta}{}^{\zeta}{}_{\delta}{}^{\iota} R_{\gamma \zeta \epsilon \iota} + C_{8} H_{\alpha \beta}{}^{\delta} H^{\alpha \beta \gamma} H_{\epsilon \zeta}{}^{\kappa} H^{\epsilon \zeta \iota} R_{\gamma \iota \delta \kappa}\nn\\&& + C_{9} H^{\alpha \beta \gamma} H^{\delta \epsilon \zeta} R_{\alpha \beta \delta}{}^{\iota} R_{\gamma \iota \epsilon \zeta}  + C_{10} H_{\alpha}{}^{\delta \epsilon} H^{\alpha \beta \gamma} R_{\beta}{}^{\zeta}{}_{\delta}{}^{\iota} R_{\gamma \iota \epsilon \zeta} + C_{11} H_{\alpha}{}^{\delta \epsilon} H^{\alpha \beta \gamma} R_{\beta}{}^{\zeta}{}_{\gamma}{}^{\iota} R_{\delta \zeta \epsilon \iota}\nn\\&& + C_{12} H_{\alpha \beta}{}^{\delta} H^{\alpha \beta \gamma} R_{\gamma}{}^{\epsilon \zeta \iota} R_{\delta \zeta \epsilon \iota}  + C_{13} H_{\alpha \beta}{}^{\delta} H^{\alpha \beta \gamma} H_{\gamma}{}^{\epsilon \zeta} H_{\epsilon}{}^{\iota \kappa} R_{\delta \iota \zeta \kappa} + C_{14} H_{\alpha}{}^{\delta \epsilon} H^{\alpha \beta \gamma} H_{\beta \delta}{}^{\zeta} H_{\gamma}{}^{\iota \kappa} R_{\epsilon \iota \zeta \kappa}\nn\\&& + C_{15} H_{\alpha \beta}{}^{\delta} H^{\alpha \beta \gamma} H_{\gamma}{}^{\epsilon \zeta} H_{\delta}{}^{\iota \kappa} R_{\epsilon \iota \zeta \kappa}  + C_{16} H_{\alpha}{}^{\delta \epsilon} H^{\alpha \beta \gamma} \nabla_{\iota}H_{\delta \epsilon \zeta} \nabla^{\iota}H_{\beta \gamma}{}^{\zeta}\nn\\&& + C_{17} H_{\alpha}{}^{\delta \epsilon} H^{\alpha \beta \gamma} \nabla_{\zeta}H_{\gamma \epsilon \iota} \nabla^{\iota}H_{\beta \delta}{}^{\zeta} + C_{18} H_{\alpha}{}^{\delta \epsilon} H^{\alpha \beta \gamma} \nabla_{\iota}H_{\gamma \epsilon \zeta} \nabla^{\iota}H_{\beta \delta}{}^{\zeta}\nn\\&& + C_{19} H_{\alpha \beta}{}^{\delta} H^{\alpha \beta \gamma} \nabla_{\zeta}H_{\delta \epsilon \iota} \nabla^{\iota}H_{\gamma}{}^{\epsilon \zeta} + C_{20} H_{\alpha \beta}{}^{\delta} H^{\alpha \beta \gamma} \nabla_{\iota}H_{\delta \epsilon \zeta} \nabla^{\iota}H_{\gamma}{}^{\epsilon \zeta}\label{L21}
\end{eqnarray}
 and $\mathcal{L}_2^2$ has  the other couplings which all include  derivatives of the dilaton, \ie,
\begin{eqnarray}
\mathcal{L}_2^2&\!\!\!\!\!=\!\!\!\!\!& C_{21} H_{\beta}{}^{\epsilon \zeta} H^{\beta \gamma \delta} H_{\gamma \epsilon}{}^{\iota} H_{\delta \zeta \iota} \nabla_{\alpha}\Phi \nabla^{\alpha}\Phi + C_{22} R_{\beta \gamma \delta \epsilon} R^{\beta \gamma \delta \epsilon} \nabla_{\alpha}\Phi \nabla^{\alpha}\Phi + C_{23} H_{\beta}{}^{\epsilon \zeta} H^{\beta \gamma \delta} R_{\gamma \epsilon \delta \zeta} \nabla_{\alpha}\Phi \nabla^{\alpha}\Phi\nn\\&& + C_{24} H_{\alpha}{}^{\gamma \delta} H_{\beta}{}^{\epsilon \zeta} H_{\gamma \epsilon}{}^{\iota} H_{\delta \zeta \iota} \nabla^{\alpha}\Phi \nabla^{\beta}\Phi + C_{25} R_{\alpha}{}^{\gamma \delta \epsilon} R_{\beta \delta \gamma \epsilon} \nabla^{\alpha}\Phi \nabla^{\beta}\Phi \nn\\&&+ C_{26} H_{\alpha}{}^{\gamma \delta} H_{\beta}{}^{\epsilon \zeta} R_{\gamma \epsilon \delta \zeta} \nabla^{\alpha}\Phi \nabla^{\beta}\Phi + C_{27} H_{\gamma \delta \epsilon} H^{\gamma \delta \epsilon} \nabla^{\alpha}\Phi \nabla_{\beta}\nabla_{\alpha}\Phi \nabla^{\beta}\Phi \nn\\&&+ C_{28} H_{\alpha}{}^{\gamma \delta} H_{\beta}{}^{\epsilon \zeta} H_{\gamma \epsilon}{}^{\iota} H_{\delta \zeta \iota} \nabla^{\beta}\nabla^{\alpha}\Phi + C_{29} H_{\alpha}{}^{\gamma \delta} H_{\beta}{}^{\epsilon \zeta} H_{\gamma \delta}{}^{\iota} H_{\epsilon \zeta \iota} \nabla^{\beta}\nabla^{\alpha}\Phi \nn\\&&+ C_{30} H_{\alpha}{}^{\gamma \delta} H_{\beta \gamma}{}^{\epsilon} H_{\delta}{}^{\zeta \iota} H_{\epsilon \zeta \iota} \nabla^{\beta}\nabla^{\alpha}\Phi + C_{31} H_{\gamma \delta}{}^{\zeta} H^{\gamma \delta \epsilon} R_{\alpha \epsilon \beta \zeta} \nabla^{\beta}\nabla^{\alpha}\Phi + C_{32} R_{\alpha}{}^{\gamma \delta \epsilon} R_{\beta \delta \gamma \epsilon} \nabla^{\beta}\nabla^{\alpha}\Phi \nn\\&&+ C_{33} H_{\alpha}{}^{\gamma \delta} H_{\gamma}{}^{\epsilon \zeta} R_{\beta \epsilon \delta \zeta} \nabla^{\beta}\nabla^{\alpha}\Phi + C_{34} H_{\alpha}{}^{\gamma \delta} H_{\beta}{}^{\epsilon \zeta} R_{\gamma \epsilon \delta \zeta} \nabla^{\beta}\nabla^{\alpha}\Phi + C_{35} H_{\alpha}{}^{\delta \epsilon} \nabla^{\alpha}\Phi \nabla^{\beta}\Phi \nabla_{\gamma}H_{\beta \delta \epsilon} \nabla^{\gamma}\Phi \nn\\&&+ C_{36} \nabla_{\alpha}\Phi \nabla^{\alpha}\Phi \nabla_{\beta}\Phi \nabla^{\beta}\Phi \nabla_{\gamma}\Phi \nabla^{\gamma}\Phi + C_{37} \nabla_{\alpha}\Phi \nabla^{\alpha}\Phi \nabla^{\beta}\Phi \nabla_{\gamma}\nabla_{\beta}\Phi \nabla^{\gamma}\Phi \nn\\&&+ C_{38} H_{\beta}{}^{\delta \epsilon} H_{\gamma \delta \epsilon} \nabla^{\alpha}\Phi \nabla^{\beta}\Phi \nabla^{\gamma}\nabla_{\alpha}\Phi + C_{39} H_{\beta}{}^{\delta \epsilon} H_{\gamma \delta \epsilon} \nabla^{\beta}\nabla^{\alpha}\Phi \nabla^{\gamma}\nabla_{\alpha}\Phi\nn\\&& + C_{40} \nabla^{\alpha}\Phi \nabla^{\beta}\Phi \nabla_{\gamma}\nabla_{\beta}\Phi \nabla^{\gamma}\nabla_{\alpha}\Phi + C_{41} \nabla^{\beta}\nabla^{\alpha}\Phi \nabla_{\gamma}\nabla_{\beta}\Phi \nabla^{\gamma}\nabla_{\alpha}\Phi \nn\\&&+ C_{42} H_{\beta}{}^{\delta \epsilon} H_{\gamma \delta \epsilon} \nabla_{\alpha}\Phi \nabla^{\alpha}\Phi \nabla^{\gamma}\nabla^{\beta}\Phi + C_{43} H_{\alpha}{}^{\delta \epsilon} \nabla^{\alpha}\Phi \nabla_{\gamma}H_{\beta \delta \epsilon} \nabla^{\gamma}\nabla^{\beta}\Phi \nn\\&&+ C_{44} \nabla_{\alpha}\Phi \nabla^{\alpha}\Phi \nabla_{\gamma}\nabla_{\beta}\Phi \nabla^{\gamma}\nabla^{\beta}\Phi + C_{45} H_{\alpha \gamma}{}^{\epsilon} H_{\beta \delta \epsilon} \nabla^{\alpha}\Phi \nabla^{\beta}\Phi \nabla^{\delta}\nabla^{\gamma}\Phi \nn\\&&+ C_{46} R_{\alpha \gamma \beta \delta} \nabla^{\alpha}\Phi \nabla^{\beta}\Phi \nabla^{\delta}\nabla^{\gamma}\Phi + C_{47} H_{\alpha \gamma}{}^{\epsilon} H_{\beta \delta \epsilon} \nabla^{\beta}\nabla^{\alpha}\Phi \nabla^{\delta}\nabla^{\gamma}\Phi + C_{48} R_{\alpha \gamma \beta \delta} \nabla^{\beta}\nabla^{\alpha}\Phi \nabla^{\delta}\nabla^{\gamma}\Phi \nn\\&&+ C_{49} H_{\beta}{}^{\delta \epsilon} \nabla^{\alpha}\Phi \nabla^{\gamma}\nabla^{\beta}\Phi \nabla_{\epsilon}H_{\alpha \gamma \delta} + C_{50} H^{\gamma \delta \epsilon} \nabla^{\alpha}\Phi \nabla^{\beta}\nabla_{\alpha}\Phi \nabla_{\epsilon}H_{\beta \gamma \delta} \nn\\&&+ C_{51} \nabla^{\alpha}\Phi \nabla^{\beta}\Phi \nabla_{\delta}H_{\beta \gamma \epsilon} \nabla^{\epsilon}H_{\alpha}{}^{\gamma \delta} + C_{52} \nabla^{\beta}\nabla^{\alpha}\Phi \nabla_{\delta}H_{\beta \gamma \epsilon} \nabla^{\epsilon}H_{\alpha}{}^{\gamma \delta} \nn\\&&+ C_{53} \nabla^{\alpha}\Phi \nabla^{\beta}\Phi \nabla_{\epsilon}H_{\beta \gamma \delta} \nabla^{\epsilon}H_{\alpha}{}^{\gamma \delta} + C_{54} \nabla^{\beta}\nabla^{\alpha}\Phi \nabla_{\epsilon}H_{\beta \gamma \delta} \nabla^{\epsilon}H_{\alpha}{}^{\gamma \delta} \nn\\&&+ C_{55} \nabla_{\alpha}\Phi \nabla^{\alpha}\Phi \nabla_{\epsilon}H_{\beta \gamma \delta} \nabla^{\epsilon}H^{\beta \gamma \delta} + C_{56} H_{\alpha}{}^{\beta \gamma} R_{\gamma \zeta \delta \epsilon} \nabla^{\alpha}\Phi \nabla^{\zeta}H_{\beta}{}^{\delta \epsilon} \nn\\&&+ C_{57} H_{\beta \gamma}{}^{\epsilon} H^{\beta \gamma \delta} H_{\delta}{}^{\zeta \iota} \nabla^{\alpha}\Phi \nabla_{\iota}H_{\alpha \epsilon \zeta} + C_{58} H_{\alpha}{}^{\beta \gamma} H_{\delta \epsilon}{}^{\iota} H^{\delta \epsilon \zeta} \nabla^{\alpha}\Phi \nabla_{\iota}H_{\beta \gamma \zeta} \nn\\&&+ C_{59} H_{\alpha}{}^{\beta \gamma} H_{\beta}{}^{\delta \epsilon} H_{\delta}{}^{\zeta \iota} \nabla^{\alpha}\Phi \nabla_{\iota}H_{\gamma \epsilon \zeta} + C_{60} H_{\alpha}{}^{\beta \gamma} H_{\beta}{}^{\delta \epsilon} H_{\gamma}{}^{\zeta \iota} \nabla^{\alpha}\Phi \nabla_{\iota}H_{\delta \epsilon \zeta}\label{AA}
\end{eqnarray}
 We have  also dropped the prime on the parameters and relabelled them from 1 to 60. We have found the minimum number of couplings which have no dilaton, \ie, $\mathcal{L}_2^1$,  by solving  the equation \eqref{LL2} for constant dilaton which produces 20 relations between $\delta C$s. 

In the second scheme, we are going to write all 60 couplings such that there is no dilaton in any of them.    We have found that there are many such schemes. We choose the following scheme:  
\begin{eqnarray}
\mathcal{L}'_2&=&\mathcal{L}^1_2+\mathcal{L}^3_2\label{L2123}
\end{eqnarray}
where $\mathcal{L}^1_2$ is the same as \eqref{L21} and  $\mathcal{L}^3_2$ contains  the following 40 couplings:
\begin{eqnarray}
\mathcal{L}^3_2\!\!=\!\!\!\!\!\!\!\!\!&&  C_{21} H_{\alpha \beta \gamma} H^{\alpha \beta \gamma} H_{\delta \epsilon}{}^{\iota} H^{\delta \epsilon \zeta} H_{\zeta}{}^{\kappa \mu} H_{\iota \kappa \mu} + C_{22} H_{\alpha \beta \gamma} H^{\alpha \beta \gamma} H_{\delta \epsilon \zeta} H^{\delta \epsilon \zeta} H_{\iota \kappa \mu} H^{\iota \kappa \mu} \nn\\&&+ C_{23} H_{\alpha}{}^{\gamma \delta} H_{\beta}{}^{\epsilon \zeta} H_{\gamma \epsilon}{}^{\iota} H_{\delta \zeta \iota} R^{\alpha \beta} + C_{24} H_{\alpha}{}^{\gamma \delta} H_{\beta \gamma \delta} H_{\epsilon \zeta \iota} H^{\epsilon \zeta \iota} R^{\alpha \beta} + C_{25} R_{\alpha}{}^{\gamma} R^{\alpha \beta} R_{\beta \gamma}\nn\\&& + C_{26} H_{\alpha}{}^{\delta \epsilon} H^{\alpha \beta \gamma} H_{\beta \delta}{}^{\zeta} H_{\gamma \epsilon \zeta} R + C_{27} H_{\alpha \beta}{}^{\delta} H^{\alpha \beta \gamma} H_{\gamma}{}^{\epsilon \zeta} H_{\delta \epsilon \zeta} R + C_{28} H_{\alpha \beta \gamma} H^{\alpha \beta \gamma} H_{\delta \epsilon \zeta} H^{\delta \epsilon \zeta} R \nn\\&&+ C_{29} R_{\alpha \beta} R^{\alpha \beta} R + C_{30} H_{\alpha \beta \gamma} H^{\alpha \beta \gamma} R^2 + C_{31} R^3 + C_{32} R R_{\alpha \gamma \beta \delta} R^{\alpha \beta \gamma \delta} + C_{33} H_{\alpha}{}^{\delta \epsilon} H^{\alpha \beta \gamma} R R_{\beta \delta \gamma \epsilon} \nn\\&&+ C_{34} H_{\alpha \beta \gamma} H^{\alpha \beta \gamma} R_{\delta \zeta \epsilon \iota} R^{\delta \epsilon \zeta \iota} + C_{35} H_{\alpha \beta \gamma} H^{\alpha \beta \gamma} H_{\delta}{}^{\iota \kappa} H^{\delta \epsilon \zeta} R_{\epsilon \iota \zeta \kappa} + C_{36} H^{\gamma \delta \epsilon} R^{\alpha \beta} \nabla_{\beta}\nabla_{\epsilon}H_{\alpha \gamma \delta}\nn\\&& + C_{37} H_{\alpha}{}^{\gamma \delta} R^{\alpha \beta} \nabla_{\beta}\nabla_{\epsilon}H_{\gamma \delta}{}^{\epsilon} + C_{38} \nabla_{\alpha}R^{\alpha \beta} \nabla_{\gamma}R_{\beta}{}^{\gamma} + C_{39} H^{\alpha \beta \gamma} R \nabla_{\gamma}\nabla_{\delta}H_{\alpha \beta}{}^{\delta}\label{RH}\\&& + C_{40} H^{\alpha \beta \gamma} R \nabla_{\delta}\nabla^{\delta}H_{\alpha \beta \gamma} + C_{41} H_{\alpha}{}^{\gamma \delta} R^{\alpha \beta} \nabla_{\delta}\nabla_{\epsilon}H_{\beta \gamma}{}^{\epsilon} + C_{42} H^{\alpha \beta \gamma} R_{\beta \epsilon \gamma \zeta} \nabla_{\delta}\nabla^{\zeta}H_{\alpha}{}^{\delta \epsilon}\nn\\&& + C_{43} H_{\alpha \beta}{}^{\delta} H^{\alpha \beta \gamma} H^{\epsilon \zeta \iota} \nabla_{\delta}\nabla_{\iota}H_{\gamma \epsilon \zeta} + C_{44} H_{\alpha \beta}{}^{\delta} H^{\alpha \beta \gamma} H_{\gamma}{}^{\epsilon \zeta} \nabla_{\delta}\nabla_{\iota}H_{\epsilon \zeta}{}^{\iota} + C_{45} H^{\gamma \delta \epsilon} \nabla_{\alpha}R^{\alpha \beta} \nabla_{\epsilon}H_{\beta \gamma \delta}\nn\\&& + C_{46} \nabla_{\alpha}R^{\alpha \beta \gamma \delta} \nabla_{\epsilon}R_{\beta}{}^{\epsilon}{}_{\gamma \delta} + C_{47} H^{\gamma \delta \epsilon} R^{\alpha \beta} \nabla_{\epsilon}\nabla_{\beta}H_{\alpha \gamma \delta} + C_{48} \nabla_{\alpha}H^{\alpha \beta \gamma} \nabla_{\epsilon}\nabla_{\gamma}\nabla_{\delta}H_{\beta}{}^{\delta \epsilon} \nn\\&&+ C_{49} H_{\alpha}{}^{\gamma \delta} R^{\alpha \beta} \nabla_{\epsilon}\nabla_{\delta}H_{\beta \gamma}{}^{\epsilon} + C_{50} \nabla^{\delta}\nabla_{\alpha}H^{\alpha \beta \gamma} \nabla_{\epsilon}\nabla_{\delta}H_{\beta \gamma}{}^{\epsilon} + C_{51} \nabla_{\alpha}H^{\alpha \beta \gamma} \nabla_{\epsilon}\nabla_{\delta}\nabla^{\epsilon}H_{\beta \gamma}{}^{\delta}\nn\\&& + C_{52} H_{\alpha}{}^{\gamma \delta} R^{\alpha \beta} \nabla_{\epsilon}\nabla^{\epsilon}H_{\beta \gamma \delta} + C_{53} \nabla^{\delta}\nabla_{\alpha}H^{\alpha \beta \gamma} \nabla_{\epsilon}\nabla^{\epsilon}H_{\beta \gamma \delta} + C_{54} \nabla_{\alpha}H^{\alpha \beta \gamma} \nabla_{\epsilon}\nabla^{\epsilon}\nabla_{\delta}H_{\beta \gamma}{}^{\delta}\nn\\&& + C_{55} H^{\alpha \beta \gamma} R_{\beta \delta \gamma \epsilon} \nabla_{\zeta}\nabla^{\zeta}H_{\alpha}{}^{\delta \epsilon} + C_{56} H_{\alpha}{}^{\delta \epsilon} H^{\alpha \beta \gamma} H_{\beta \delta}{}^{\zeta} \nabla_{\zeta}\nabla_{\iota}H_{\gamma \epsilon}{}^{\iota} + C_{57} H^{\alpha \beta \gamma} R_{\beta \epsilon \gamma \zeta} \nabla^{\zeta}\nabla_{\delta}H_{\alpha}{}^{\delta \epsilon}\nn\\&& + C_{58} H_{\alpha}{}^{\delta \epsilon} H^{\alpha \beta \gamma} \nabla_{\zeta}H_{\beta \delta}{}^{\zeta} \nabla_{\iota}H_{\gamma \epsilon}{}^{\iota} + C_{59} H_{\alpha \beta}{}^{\delta} H^{\alpha \beta \gamma} \nabla_{\epsilon}H_{\gamma}{}^{\epsilon \zeta} \nabla_{\iota}H_{\delta \zeta}{}^{\iota} + C_{60} H_{\alpha}{}^{\delta \epsilon} H^{\alpha \beta \gamma} \nabla^{\zeta}H_{\beta \gamma \delta} \nabla_{\iota}H_{\epsilon \zeta}{}^{\iota}\nn
\end{eqnarray}
When metric and $B$-field are constant, there is no relation between  $\delta C$s, hence, the minimum number of couplings between only dilaton is zero. On the other hand, when metric and $B$-field are constant, there is no coupling in \eqref{L2123}.

The parameters in the field redefinitions that change the action \eqref{eq.final0} to \eqref{L2122} or \eqref{L2123}, are functions of $\delta C_1,\cdots, \delta C_{705}$ and some of unfixed parameters $b_i$s, $c_i$s and $d_i$s in \eqref{eq.g.1}. In the above schemes, $\delta C_i=g_i(C_1,\cdots, C_{705})$ where $i$ is 60 numbers among $1,\cdots, 705$  depending on the scheme, and all others are $\delta C_j=-C_j$. The unfixed parameters $b_i$s, $c_i$s and $d_i$s  satisfy the following relation:
\begin{align}
 \mathcal{J}_2+\mathcal{K}_2=0\label{JK2}
\end{align}
One may ask if there are similar relations as \eqref{PgB1} for the residual field redefinitions at    order $\alpha'^2$. To answer this question, we  write the above equation in the local frame and then solve the resulting  equations to find some relations between the  unfixed parameters $b_i$s, $c_i$s and $d_i$s and the arbitrary parameters in the total derivative terms, \ie,  $J_i$s.  Then one can use the  arbitrary  parameters  $J_i$ to cancel the field redefinitions which involve terms with more than two derivatives. The resulting field redefinitions which we call them $\delta\hat{\Phi}^{(2)}, \delta\hat{g}_{\mu\nu}^{(2)}, \delta\hat{B}_{\mu\nu}^{(2)}$  can then be rewritten as 
\begin{eqnarray}
\int d^Dx\sqrt{-g}e^{-2\Phi}\delta\hat{\Phi}^{(2)}&=&\frac{\delta S_0}{\delta g_{\mu\nu}}\delta \bar{g}_{\mu\nu}^{(1)}+\frac{\delta S_0}{\delta B_{\mu\nu}}\delta \bar{B}_{\mu\nu}^{(1)}+\frac{\delta S_0}{\delta \Phi}\delta \bar{\Phi}^{(1)}+\frac{\delta S_0}{\delta g_{\mu\nu}}g_{\mu\nu}\delta \bar{\Phi_1}^{(1)}\nn\\
\int d^Dx\sqrt{-g}e^{-2\Phi}\delta\hat{g}_{\mu\nu}^{(2)}&=&\frac{\delta S_0}{\delta g^{\mu\nu}}\delta \bar{\Phi_2}^{(1)} +\frac{\delta S_0}{\delta g_{\alpha\beta}}g_{\alpha\beta}\delta \bar{g_1}_{\mu\nu}^{(1)}+\frac{\delta S_0}{\delta g^{\alpha\{\mu}}\delta \bar{g_2}^{(1)}_{\nu\}\alpha}+\frac{\delta S_0}{\delta g_{\alpha\beta}}g_{\alpha\beta}\delta \bar{g_3}_{\mu\nu}^{(1)}\nn\\&&+\frac{\delta S_0}{\delta \Phi}\delta \bar{g_4}_{\mu\nu}^{(1)}+\frac{\delta S_0}{\delta B^{\alpha\{\mu}}\delta \bar{B_1}^{(1)}_{\nu\}\alpha} +\frac{\delta S_0}{\delta B_{\alpha\beta}}\delta \bar{B_2}^{(1)}_{\alpha\beta\mu\nu}\nn\\
\int d^Dx\sqrt{-g}e^{-2\Phi}\delta\hat{B}_{\mu\nu}^{(2)}&=&\frac{\delta S_0}{\delta B^{\alpha[\mu}}\delta \bar{g_5}^{(1)}_{\nu]\alpha} +\frac{\delta S_0}{\delta B_{\mu\nu}}\delta \bar{\Phi_3}^{(1)}+\frac{\delta S_0}{\delta \Phi}\delta \bar{B_3}_{\mu\nu}^{(1)}+\frac{\delta S_0}{\delta g_{\alpha\beta}}\delta \bar{B_4}^{(1)}_{\alpha\beta\mu\nu}\label{PgB2}
\end{eqnarray}
 where the tensors $\delta \bar{g}_{\mu\nu}^{(1)},\cdots\delta\bar{g_5}_{\mu\nu}^{(1)}$, $ \delta \bar{B}_{\mu\nu}^{(1)}, \cdots, \delta \bar{B_4}_{\alpha\beta\mu\nu}^{(1)}$ and $\delta \bar{\Phi}^{(1)},\cdots, \delta \bar{\Phi_3}^{(1)} $ are some specific functions of $R,H,\nabla\Phi$ and their derivatives at order $\alpha'$. Using similar steps as in \eqref{eq.field.1} and \eqref{eq.field.2}, and using the residual field redefinitions in \eqref{PgB1} and \eqref{PgB2},  one then can write the variations that the actions $S_0,S_1,S_2$ produce at order   $\alpha'^3$, up to some total derivative terms, as
\begin{align}
&\delta S_0+\delta S_1+\delta S_2=\frac{\delta S_0}{\delta g_{\alpha\beta}}\delta g^{(3)}_{\alpha\beta}+\frac{\delta S_0}{\delta B_{\alpha\beta}}\delta B^{(3)}_{\alpha\beta}+\frac{\delta S_0}{\delta \Phi}\delta \Phi^{(3)} 
\end{align}
where the deformations $\delta {g}_{\mu\nu}^{(3)}, \delta {B}_{\mu\nu}^{(3)}, \delta {\Phi}^{(3)}$  are arbitrary functions of $R,H,\nabla\Phi$ and their derivatives at order $\alpha'^3$. 

It seems similar rewriting as \eqref{PgB2} can be done for residual field redefinitions at higher orders of $\alpha'$ as well. As a result, the variations of actions $S_0,\cdots, S_{n-1}$ may produce the following contributions at order $\alpha'^n$:
\begin{align}
&\delta S_0+\cdots+\delta S_{n-1}=\frac{\delta S_0}{\delta g_{\alpha\beta}}\delta g^{(n)}_{\alpha\beta}+\frac{\delta S_0}{\delta B_{\alpha\beta}}\delta B^{(n)}_{\alpha\beta}+\frac{\delta S_0}{\delta \Phi}\delta \Phi^{(n)} \label{S0n}
\end{align}
where the deformations $\delta {g}_{\mu\nu}^{n)}, \delta {B}_{\mu\nu}^{(n)}, \delta {\Phi}^{(n)}$  are arbitrary functions of $R,H,\nabla\Phi$ and their derivatives at order $\alpha'^n$. Therefore, as long as one considers fixed couplings at orders $\alpha',\cdots,\alpha'^{n-1}$, the contributions of the field redefinitions on the actions $S_0,\cdots, S_{n-1}$ at order $\alpha'^n$ are given by \eqref{S0n}. Using high performance computer,  it is then straightforward to extend the   calculations in this paper to the order $\alpha'^n$. The minimal couplings may be written in the following scheme:
\begin{eqnarray}
\mathcal{L}_n'&=&\mathcal{L}_n^1+\mathcal{L}_n^2\label{LLL}
\end{eqnarray}
where $\mathcal{L}_n^1$ contain the minimum number of couplings for metric and $B$-field, and $\mathcal{L}_n^2$ contains all other couplings each contains derivative of dilaton. 

The minimal  independent parameters in \eqref{LLL} for $n=2$ which are given in  \eqref{L21} and \eqref{AA}  or \eqref{RH} may be calculated by    studying in details the S-matrix element of  vertex operators in string theory, or they may be found  by the T-duality constraint up to an overall factor. We leave these calculations for the future works.  

{\bf Acknowledgments}:   This work is supported by Ferdowsi University of Mashhad.

\end{document}